\RequirePackage{lineno}
\documentclass[aps,prl,twocolumn,superscriptaddress,groupedaddress,preprintnumbers]{revtex4}  
\usepackage{graphicx}  
\usepackage{dcolumn}   
\usepackage{bm}        
\usepackage{amssymb}   
\usepackage{multirow}
\usepackage{epsfig}
\usepackage{amsmath}
\usepackage[colorlinks, linkcolor=blue]{hyperref}
\usepackage{ulem}

\newcommand{\effstw}{\ensuremath{\sin^2\theta_{\text{eff}}^{\text{$\ell$}}}}

\begin{document}

\lefthyphenmin=2
\righthyphenmin=2

\widetext

\title{ Boost Asymmetry of the diboson productions in $pp$ collisions}
\affiliation{Department of Modern Physics, University of Science and Technology of China, Jinzhai Road 96, Hefei, Anhui 230026, China}
\affiliation{School of Nuclear Science and Technology, University of South China, Hengyang, Hunan 421001, China}
\affiliation{Department of Physics and Astronomy, Michigan State University, East Lansing, MI 48823, USA}

\author{Siqi Yang} \affiliation{Department of Modern Physics, University of Science and Technology of China, Jinzhai Road 96, Hefei, Anhui 230026, China}
\author{Mingzhe Xie} \affiliation{Department of Modern Physics, University of Science and Technology of China, Jinzhai Road 96, Hefei, Anhui 230026, China}
\author{Yao Fu} \affiliation{Department of Modern Physics, University of Science and Technology of China, Jinzhai Road 96, Hefei, Anhui 230026, China}
\author{Zihan Zhao} \affiliation{Department of Modern Physics, University of Science and Technology of China, Jinzhai Road 96, Hefei, Anhui 230026, China}
\author{Minghui Liu} \affiliation{Department of Modern Physics, University of Science and Technology of China, Jinzhai Road 96, Hefei, Anhui 230026, China}
\author{Liang Han} \affiliation{Department of Modern Physics, University of Science and Technology of China, Jinzhai Road 96, Hefei, Anhui 230026, China}
\author{Tie-Jiun Hou} \affiliation{School of Nuclear Science and Technology, University of South China, Hengyang, Hunan 421001, China}
\author{C.-P. Yuan} \affiliation{Department of Physics and Astronomy, Michigan State University, East Lansing, MI 48823, USA}

\begin{abstract}
We propose the boost asymmetry of the diboson productions $q_i\bar{q}_j\rightarrow VV'$ in $pp$ collisions 
($VV' = W\gamma$, $W^+W^-$ and $WZ$)
as a new experimental observable, which can provide unique information on the proton structure. 
The boost asymmetry rises as the difference in the kinematics of the two bosons, 
that are coupled to the two different quark and antiquark initial states, respectively, 
and thus reflects different features of the $q_i$ and $\bar{q}_j$ parton densities. 
By comparing 
the kinematics of the two bosons, such as the boson energy or rapidity, 
the diboson events with $\bar{q}_j$ having higher energy than 
$q_i$ can be distinguished from those with $q_i$ having higher energy than $\bar{q}_j$. 
This would provide unique information in some special parton momentum fraction regions, which 
cannot be directly proved by current $W$ and $Z$ measurements at the Large Hadron Collider 
or other deep inelastic scattering experiments.
\end{abstract}
\maketitle

\section{I. Introduction}

The vector boson productions at the Large Hadron Collider (LHC) are dominated by the 
initial state quark-antiquark ($q_i\bar{q}_j$) scattering, 
thus are highly sensitive to the 
corresponding parton densities in a large range of 
the Bjorken variable $x$, 
describing the fraction of the parton momentum to the 
energy of the proton. 
In the latest global analysis of the parton distribution functions (PDFs) such as 
the CT18, MSHT20 and NNPDF4.0, the 
single $Z$ and $W$ production rates measured at 7 and 8 TeV $pp$ collisions 
have been used and delivered 
significant impacts~\cite{CT18PDF, MSHT20, NNPDF41}.

Due to the high energy of the proton beam, the vector boson productions 
at the LHC contain 
both the $q_i(x_L)\bar{q}_j(x_S)$ contribution where the initial state quarks carry higher 
energy than the antiquarks ($x_L > x_S$), and the $\bar{q}_j(x_L)q_i(x_S)$ contribution 
where the antiquarks 
have higher energy. The ratio between 
the two cross sections, which we call as the quark exchanging fraction, 
relates to the 
parton densities of the valence $u$ and $d$ quarks in the small $x$ region, 
and the sea quarks in the relatively large $x$ region. 
An example is the 
forward backward asymmetry ($A_{FB}$) of the Drell-Yan 
$pp\rightarrow q_i\bar{q}_j \rightarrow  Z/\gamma^* \rightarrow \ell^+\ell^-$ process. 
Due to the limited knowledge of the
dilution effect, which is actually the quark exchanging fraction for the 
$u\bar{u}$ and $d\bar{d}$ cases, 
the observed $A_{FB}$ is reduced from its original value arising from the electroweak (EW)
symmetry breaking~\cite{CPCstudy}, 
and therefore induces large PDF-induced uncertainty on the determination of 
the weak mixing angle ($\effstw$), reported by the
ATLAS, CMS and LHCb measurements~\cite{ATLASstw, CMSstw, LHCbstw}. 

The quark exchanging fraction is unique information 
that is difficult to acquire from other data 
such as the Deep-Inelastic Scattering and the 
fixed-target Drell-Yan experiments. Therefore, the measurements at the LHC 
are expected to provide important input to expand our knowledge of proton structure.
However, the observed $Z$ and $W$ cross sections are always the 
mixture of the $q_i(x_L)\bar{q}_j(x_S)$ and $\bar{q}_j(x_L)q_i(x_S)$ 
contributions, 
so they could not distinguish the two initial states and fail to give direct constraint 
on the quark exchanging fraction. 
It was proposed to use $A_{FB}$ itself to constrain the dilution 
effect~\cite{ArieStudy, ATLASStudy1, ATLASStudy2}, but 
found to have large additional uncertainties due to the 
correlation with the EW $\effstw$ parameter~\cite{CPCstudy}. 

Without a direct constraint, the 
current PDF global analysis has to 
provide predictions on the quark exchanging fraction 
by combining the information of the 
LHC data and other old experimental results~\cite{OldData}. 
However, 
the combined fitting highly relies on the assumption that all the data 
should be consistent, 
while on the other hand, 
it is known that, 
e.g., the ATLAS 5 TeV, 7 TeV and 8 TeV $W$ and $Z$ 
differential cross section data have tensions with other datasets included 
in the PDF global analysis~\cite{ATLAS5TeVWZ, ATLAS7TeVWZ, ATLAS8TeVWAsym}.
Therefore, it would be important to have experimental observables which can provide 
constraint directly on the quark exchanging fraction at the LHC.

In this paper, we propose a set of new experimental observables, the 
boost asymmetries $A^{VV'}_\text{boost}$, 
in the $pp\rightarrow q_i\bar{q}_j \rightarrow VV'$ processes, 
including $VV'=W\gamma$, $W^+W^-$ and $WZ$ at the LHC. 
By observing $A^{VV'}_\text{boost}$, 
the initial states of $q_i(x_L)\bar{q}_j(x_S)$ and $\bar{q}_j(x_L)q_i(x_S)$ 
can be distinguished from each other, so that 
the quark exchanging fraction 
can be directly determined. We also perform an impact study of introducing the $A^{VV'}_\text{boost}$ 
observables to the PDF global analysis. 
It is demonstrated that the uncertainties of the relevant parton densities 
can be significantly reduced. 
~\\

\section{II. The boost asymmetry in the $VV'$ events}

\begin{figure}[!hbt]
\begin{center}
\epsfig{scale=0.2, file=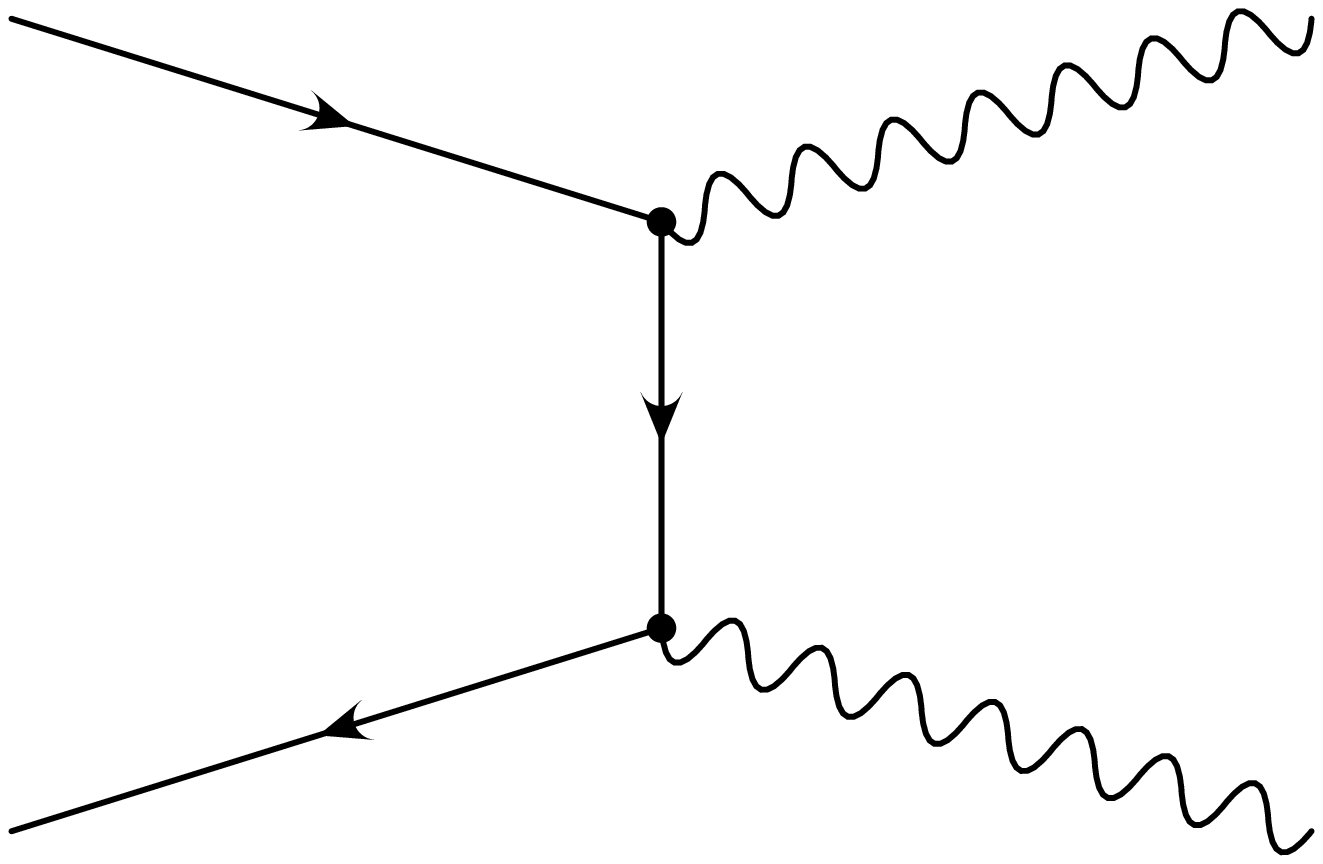}
\epsfig{scale=0.2, file=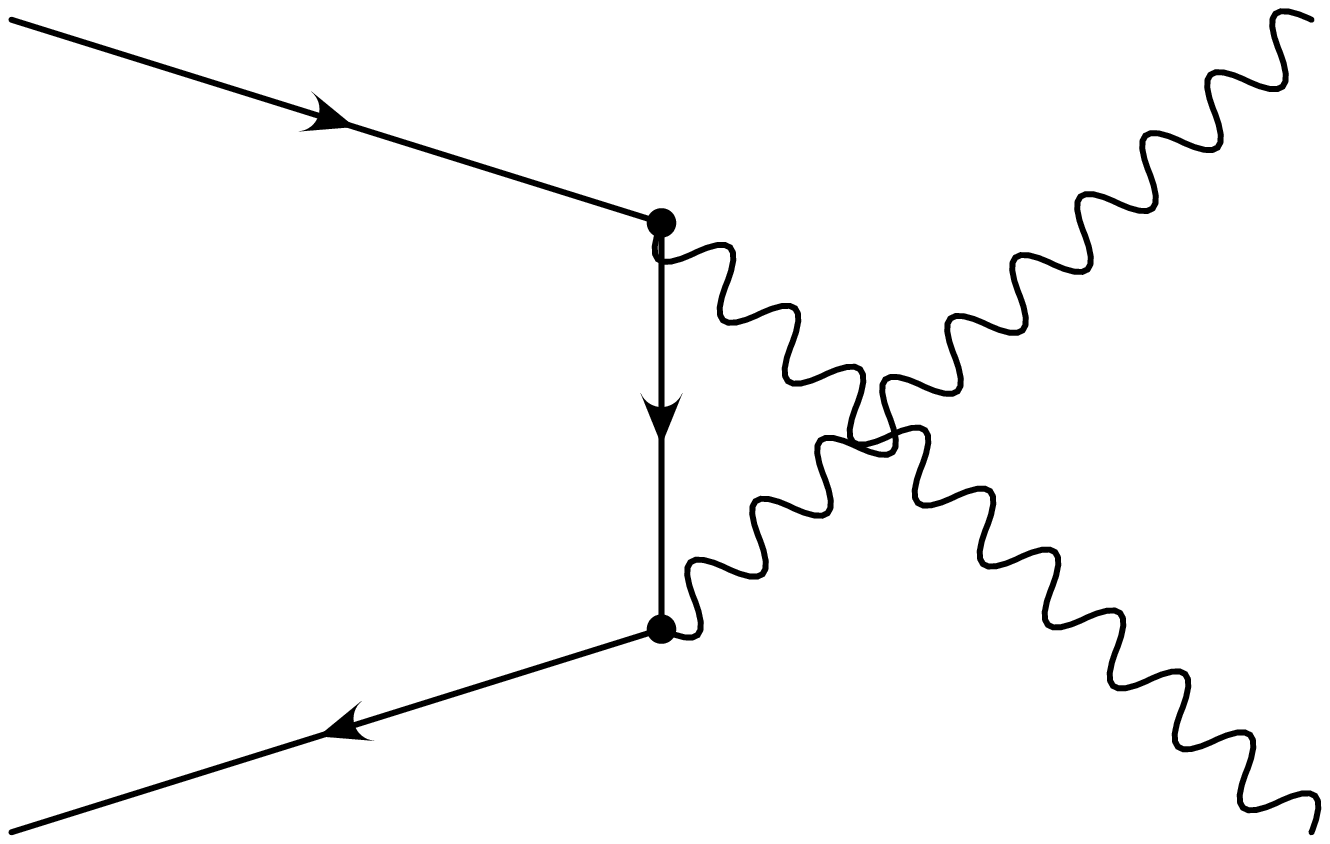}
\caption{\small The Feynman diagrams of the 
$t-$channel (left) and $u-$channel (right) processes in the $pp(q_i\bar{q}_j)\rightarrow VV'$ 
production. The $s-$channel diagram, needed for some production mode to form 
a gauge invariant set, is not shown here.}
\label{fig:feynmandiagram}
\end{center}
\end{figure}

At the LHC, the diboson events are produced dominantly by the $q_i\bar{q}_j$ initial state via 
the $t-$ and $u-$channel contributions, of which the Feynman diagrams 
are shown in Fig.~\ref{fig:feynmandiagram}. 
The two bosons are separately coupled to the quarks and antiquarks, 
and consequently 
the kinematics of the two different bosons reflect 
the energy of the corresponding quark or antiquark respectively. 
The boson kinematic can usually be represented by the rapidity ($Y$) of the boson  
or the lepton from the boson decay. 
For the $W\gamma$, $W^+W^-$ and $WZ$ processes where the two bosons 
can be experimentally distinguished, the events can be divided into 
$|Y_V| > |Y_{V'}|$ and 
$|Y_V| < |Y_{V'}|$ categories. The boost asymmetry is defined to describe 
the relative difference between the two categories:
\begin{eqnarray}
 A^{VV'}_\text{boost} = \frac{N(|Y_V| > |Y_{V'}|) - N(|Y_V| < |Y_{V'}|)}{N(|Y_V| > |Y_{V'}|) + N(|Y_V| < |Y_{V'}|)}
\end{eqnarray}
where $N$ is the number of observed events. 
Due to $q_i$ and $\bar{q}_j$ have 
different energy densities, the observation of $A_\text{boost}$ is expected to have 
non-zero values.
In the following subsections, we will 
discuss how $A^{VV'}_\text{boost}$ reflect the information of the quark exchanging fractions. 

\subsection{II-A. $A^{VV'}_\text{boost}$ in the $W\gamma$ and $WZ$ processes}

The $W^{\pm}\gamma$ production is dominated by 
the two different $u\bar{d}$ and $d\bar{u}$ initial states 
respectively, where 
$W$ and $\gamma$ can couple to either quarks or antiquarks. 
Since the $W^+\gamma$ and $W^-\gamma$ processes can be measured 
independently, it is possible to probe the different energy spectrum of $u\bar{d}$ and 
$d\bar{u}$ initial states. 
For example, according to different boson-quark-couplings ($V-q$) and 
parton densities ($q(x)$),  
the $u\bar{d}$ contribution in $W^+\gamma$ events 
can be further divided into four 
parts as: $N^{W-u(x_L)}_{\gamma-\bar{d}(x_S)}$, $N^{W-u(x_S)}_{\gamma-\bar{d}(x_L)}$, 
$N^{W-\bar{d}(x_L)}_{\gamma-u(x_S)}$ and $N^{W-\bar{d}(x_S)}_{\gamma-u(x_L)}$. 
When $u$ carries higher energy than $\bar{d}$, 
i.e. in the two $u(x_L)\bar{d}(x_S)$ cases, 
it statistically gives $|Y_\gamma |>|Y_W|$ via 
$N^{W-\bar{d}(x_S)}_{\gamma-u(x_L)}$ where $W$ couples to $\bar{d}(x_S)$ 
and $\gamma$ couples to $u(x_L)$; meanwhile, it may raise $|Y_\gamma |<|Y_W|$ via 
$N^{W-u(x_L)}_{\gamma-\bar{d}(x_S)}$, and thus partially cancel the asymmetry. However, 
the cross section of $N^{W-u(x_L)}_{\gamma-\bar{d}(x_S)}$ is smaller than 
that of $N^{W-\bar{d}(x_S)}_{\gamma-u(x_L)}$, because the $\gamma-u$ coupling 
results in larger contribution than the $\gamma-d$ coupling by a factor of about 4. 
Therefore, the cancellation on the asymmetry is not significant. 
In the same way, when $\bar{d}$ carries higher energy than $u$, i.e. in $u(x_s)\bar{d}(x_L)$ cases, 
it statistically gives $|Y_\gamma |<|Y_W|$ via 
$N^{W-\bar{d}(x_L)}_{\gamma-u(x_S)}$, 
and also would be partially cancelled by $N^{W-u(x_S)}_{\gamma-\bar{d}(x_L)}$ contribution.

Note that the relationship between $Y_W$ and $Y_\gamma$ does not exactly reflect the 
relationship between $u$ and $\bar{d}$ energy. Apart from the cancellation effect mentioned above, 
the freedom in the kinematic of internal quark propagators, 
the contributions from the $s-$channel processes 
and higher order effects, and the interference between them 
would also smear the boost asymmetry. 
In practice, the $W$ boson kinematic is usually replaced by the rapidity of the 
leptons from the decay, thus further smears $A_\text{boost}$ due to the missing neutrino. 
Nevertheless, 
these smearing effects of diminishing $A_\text{boost}$ observation are more 
related to the EW calculations which can be precisely predicted, and thus independent 
with the parton densities of the initial state quarks. As a conclusion, 
$A_\text{boost}$ in the $W^+\gamma$ events is dominated by the 
quark exchanging fraction of $u(x_L)\bar{d}(x_S)$ over $u(x_S)\bar{d}(x_L)$. 
Similarly, $A_\text{boost}$ in the $W^-\gamma$ events contains the information 
of the quark exchanging fraction of $d(x_L)\bar{u}(x_S)$ over $d(x_S)\bar{u}(x_L)$. 

To give numerical results, a
sample of $pp\rightarrow W\gamma \rightarrow \ell\nu\gamma$ is generated 
at $\sqrt{s}=13$ TeV using 
{\sc pythia} event generator~\cite{pythia}, with 700 million events in full phase space 
corresponding to about 1 ab$^{-1}$ data produced at the LHC. 
The boost asymmetry is specifically defined as:
\begin{eqnarray}
 A^{W\gamma}_\text{boost} = \frac{N(|Y_\gamma| > |Y_\ell|) - N(|Y_\gamma| < |Y_\ell|)}{N(|Y_\gamma| > |Y_\ell|) + N(|Y_\gamma| < |Y_\ell|)}
\end{eqnarray}
in terms of the rapidity of $\gamma$ and $\ell$ of the final states. 
The predicted values of  $A^{W\gamma}_\text{boost}$ in the $W^+\gamma$ and $W^-\gamma$ events 
from CT18, MSHT20 and NNPDF4.0~\cite{CT18PDF, MSHT20, NNPDF41}, together with 
the corresponding PDF uncertainties are summarized in Table~\ref{tab:WgammaResults}. 

\begin{table}[hbt]
\begin{center}
\begin{tabular}{l|c|c}
\hline \hline
  & $W^+\gamma$ events & $W^-\gamma$ events \\
\hline
$A^{W\gamma}_\text{boost}$ prediction  & 0.457& -0.119   \\
in CT18 & $\pm$0.006(PDF) &$\pm$0.008(PDF) \\
\hline
$A^{W\gamma}_\text{boost}$ prediction  & 0.446&-0.114 \\
in MSHT20 & $\pm$0.004(PDF) &$\pm$0.005(PDF) \\
\hline
$A^{W\gamma}_\text{boost}$ prediction  & 0.444&-0.111  \\
in NNPDF4.0 & $\pm$0.004(PDF) & $\pm$0.004(PDF) \\
\hline \hline
\end{tabular}
\caption{\small The boost asymmetry and the corresponding PDF-induced uncertainties predicted 
from CT18, MSHT20 and NNPDF4.0, in the $W^+\gamma\rightarrow \ell^+\nu\gamma$ and 
$W^-\gamma\rightarrow \ell^-\nu\gamma$ events. PDF uncertainties correspond to $68\%$ C.L.}
\label{tab:WgammaResults}
\end{center}
\end{table} 

The asymmetry $A^{W\gamma}_\text{boost}$ in the $W^+\gamma$ event has 
a positive large value, namely $\gamma$ is more boosted 
than the $W$ boson, because of three reasons. Firstly, 
since the probability of the valence $u$ quark having higher 
energy is greater than that of the sea $\bar{d}$ 
quark, 
$N^{W-\bar{d}(x_S)}_{\gamma-u(x_L)}$ which 
gives $|Y_\gamma| > |Y_\ell|$, 
would surpass $N^{W-\bar{d}(x_L)}_{\gamma-u(x_S)}$ which gives $|Y_\gamma| < |Y_\ell|$. 
Secondly, the cancellation of $N^{W-u(x_L)}_{\gamma-\bar{d}(x_S)}$ 
to the dominating $N^{W-\bar{d}(x_S)}_{\gamma-u(x_L)}$ 
is suppressed due to the charge-determined $\gamma-q$ couplings. 
Thirdly, 
the massless $\gamma$ would be more boosted than the massive $W$ boson. 
As a result, the boost asymmetry is enhanced. 

On the contrary, 
$A^{W\gamma}_\text{boost}$ in the $W^-\gamma$ event has a 
smaller negative value. In this $\bar{u}d$ process, $\gamma$ can acquire higher energy 
by coupling to the valence 
$d$ quark, but 
the dominating contribution $N^{W-\bar{u}(x_S)}_{\gamma-d(x_L)}$ 
is suppressed by the $\gamma$-$d$-$\bar{d}$ vertex. The $W$ boson can 
acquire higher energy 
via $N^{W-d(x_L)}_{\gamma-\bar{u}(x_S)}$, but its boost is limited due to its heavy mass. 
Consequently, neither $W$ nor $\gamma$ could significantly lead the boost 
after the cancellation.

Similarly, $A_\text{boost}$ can be defined in the $WZ$ events, and is also 
sensitive to the quark exchanging fraction of $u\bar{d}$ and $d\bar{u}$. Due 
to the smaller cross section and full lepton decay branching ratios of the 
$W$ and $Z$ boson, the boost asymmetry in the $WZ$ processes is less significant 
than that in the $W\gamma$ ones. However, the $A^{WZ}_\text{boost}$ 
observation would be more feasible than $A^{W\gamma}_\text{boost}$ 
at the LHCb, where the precise photon measurement is not practicable. 
Thus the LHCb $A^{WZ}_\text{boost}$ observation may provide complementary information 
in a much forward phase space, which cannot be covered by the acceptance of the 
ATLAS and CMS detectors.
~\\

\subsection{II-B. $A_\text{boost}$ in the $W^+W^-$ process}

The $W^+W^-$ process is dominated by the $u\bar{u}$ and $d\bar{d}$ initial states, 
and has different sensitivities to 
the quark exchanging fraction from the 
$W\gamma$ and $WZ$ events. The $W^+$ boson is always coupled 
to the positive charged $u$ or $\bar{d}$ quark, while the $W^-$ boson is  
coupled to the negative charged $\bar{u}$ or $d$ quark. Therefore, there is 
no cancellation due to the exchange of the boson-to-quark couplings in the $W^+W^-$ event.
Instead, the cancellation rises between the $u\bar{u}$ and $d\bar{d}$ contributions.
In the $u\bar{u}$ subprocess, 
the large cross section part
$N^{W^+-u(x_L)}_{W^--\bar{u}(x_S)}$ 
statistically gives $W^+$ the higher energy, while 
the higher energy $W^-$ events are mainly produced by the small cross section 
$N^{W^+-u(x_S)}_{W^--\bar{u}(x_L)}$ contribution, 
so that $W^+$ leads the boost. On the contrary, in the $d\bar{d}$ 
subprocess, it is the large cross section$N^{W^--d(x_L)}_{W^+-\bar{d}(x_S)}$ 
part 
gives $W^-$ the higher energy, 
while the small cross section 
$N^{W^--d(x_S)}_{W^+-\bar{d}(x_L)}$ 
has $W^+$ the higher energy, so that $W^-$ leads the boost.
Since the overall cross section and 
boost kinematics of $u\bar{u}$ differ from those of $d\bar{d}$, the observed boost 
asymmetry in the $W^+W^-$ is expected to be non-zero, 
and can be defined 
in terms of the rapidity of the leptons of the $W$ boson decay in the final state as:
\begin{eqnarray}
 A^{WW}_\text{boost} = \frac{N(|Y_{\ell^+}|>|Y_{\ell^-}|) - N(|Y_{\ell^+}|<|Y_{\ell^-}|)}{N(|Y_{\ell^+}|>|Y_{\ell^-}|) + N(|Y_{\ell^+}|<|Y_{\ell^-}|)}.
\end{eqnarray}
The boost asymmetry in the $W^+W^-$ event was previously discussed in Ref.~\cite{WWprevious}, 
of which it was expected to have some sensitivity in new physics search. 
In this work, we will 
demonstrate that $A^{WW}_\text{boost}$ is a useful 
observable in the PDF global analysis. 
The numerical predictions of {\sc pythia} with various PDFs are listed in Table~\ref{tab:WWResults}. 

\begin{table}[hbt]
\begin{center}
\begin{tabular}{l|c}
\hline \hline
$A^{WW}_\text{boost}$ prediction in CT18 & -0.051$\pm$0.005(PDF)    \\
\hline
$A^{WW}_\text{boost}$ prediction in MSHT20 & -0.051$\pm$0.004(PDF)   \\
\hline
$A^{WW}_\text{boost}$ prediction in NNPDF4.0 & -0.051$\pm$0.002(PDF)   \\
\hline \hline
\end{tabular}
\caption{\small The boost asymmetry and the corresponding PDF-induced uncertainties predicted  in the $W^+W^-\rightarrow \ell^+\nu\ell^-\nu$ event 
at the 13 TeV LHC, predicted from CT18, MSHT20 and NNPDF4.0. 
PDF uncertainties correspond to $68\%$ C.L.}
\label{tab:WWResults}
\end{center}
\end{table} 

Finally, we would like to discuss the boost asymmetry in the $Z\gamma$ production. 
It is also dominated by 
$u\bar{u}$ and $d\bar{d}$ initial states, thus could have the same sensitivities to the 
quark exchanging fractions as the $W^+W^-$ events. However, the sensitivity is almost 
completely cancelled due to the exchange of the boson-quark-coupling, 
namely the interchanging possibilities of either higher energy $q(x_L)$ or $\bar{q}(x_L)$ 
radiating a photon or $Z$ boson are equal. 
Even though there is still a sizable asymmetry, as shown in Table~\ref{tab:ZgammaResults}, 
such asymmetry is purely raised by the mass difference between $Z$ and $\gamma$ and 
has very little sensitivity to the quark densities. 
This is also indicated by the much smaller PDF uncertainties in Table~\ref{tab:ZgammaResults} 
than those of $A^{W\gamma}_\text{boost}$ and $A^{WW}_\text{boost}$. 
Therefore, for the $Z\gamma$ production at the LHC, 
even though it has relatively larger cross section than $W^+W^-$ 
and is easy to be measured, it would not provide as large PDF-sensitivity boost 
asymmetry as in the $W\gamma$ and $WW$ events.

\begin{table}[hbt]
\begin{center}
\begin{tabular}{l|c}
\hline \hline
$A^{Z\gamma}_\text{boost}$ prediction in CT18 & 0.124$\pm$0.0008(PDF)    \\
\hline
$A^{Z\gamma}_\text{boost}$ prediction in MSHT20 & 0.124$\pm$0.0005(PDF)   \\
\hline
$A^{Z\gamma}_\text{boost}$ prediction in NNPDF4.0 & 0.126$\pm$0.0003(PDF)   \\
\hline \hline
\end{tabular}
\caption{\small The boost asymmetry and the corresponding PDF-induced uncertainties 
in the $Z\gamma$ event at the 13 TeV LHC, predicted by CT18, MSHT20 and 
NNPDF4.0. PDF 
uncertainties correspond to $68\%$ C.L.}
\label{tab:ZgammaResults}
\end{center}
\end{table} 

\section{III. Impact study of introducing $A^{VV'}_\text{boost}$ into the PDF global fitting}

In this section, we present an impact study of introducing $A^{VV'}_\text{boost}$ into 
the PDF global fitting, by using pseudo-data samples of $W\gamma$ and $WW$ events 
corresponding to 1 ab$^{-1}$ data at the 13 TeV LHC as new experimental input. 
The error PDF Updating Method Package 
(ePump)~\cite{epump} is used which can efficiently update the PDF with a new data input 
in the way equivalent to the PDF global fitting based on the Hessian approach.
The central and error set predictions of CT18 PDFs on $A^{W\gamma}_\text{boost}$ and 
$A^{WW}_\text{boost}$ are used as inputs to ePump.

\begin{figure}[!hbt]
\begin{center}
\epsfig{scale=0.3, file=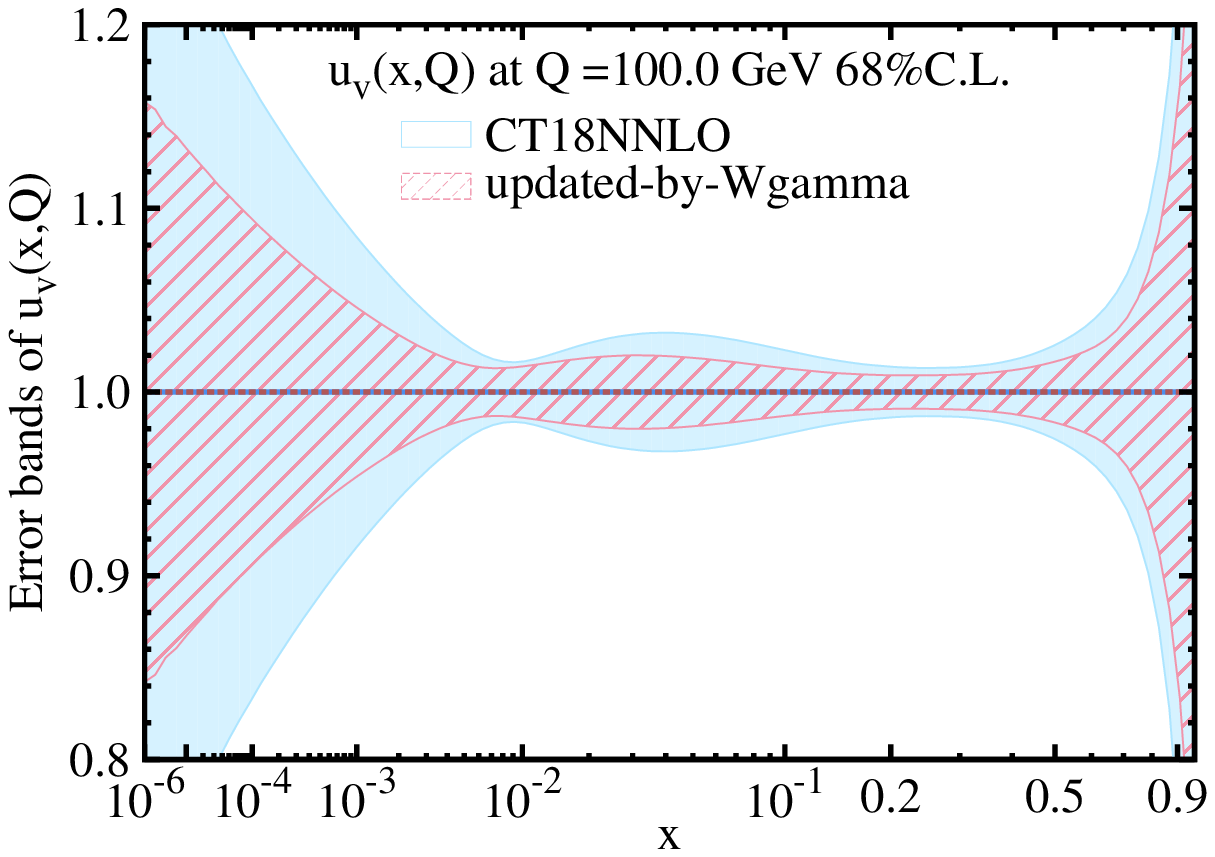}
\epsfig{scale=0.3, file=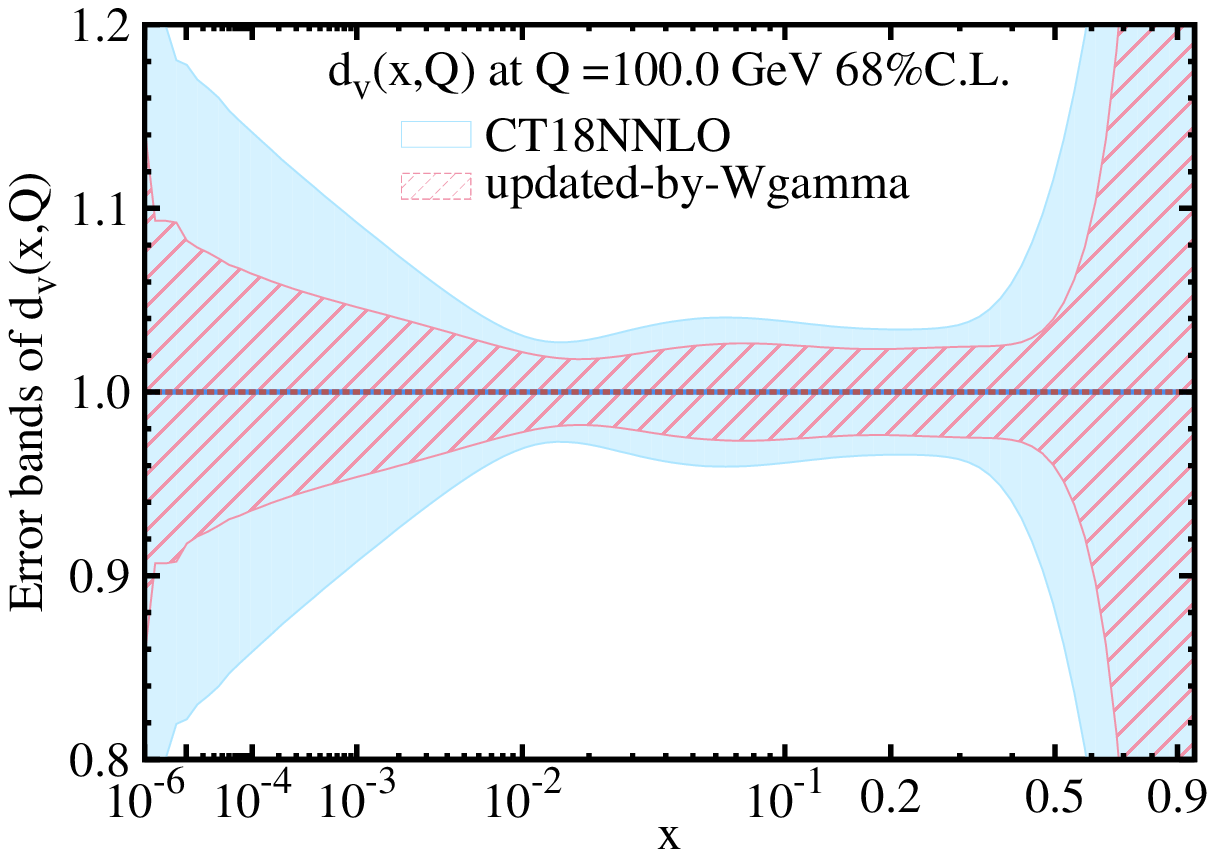}
\epsfig{scale=0.3, file=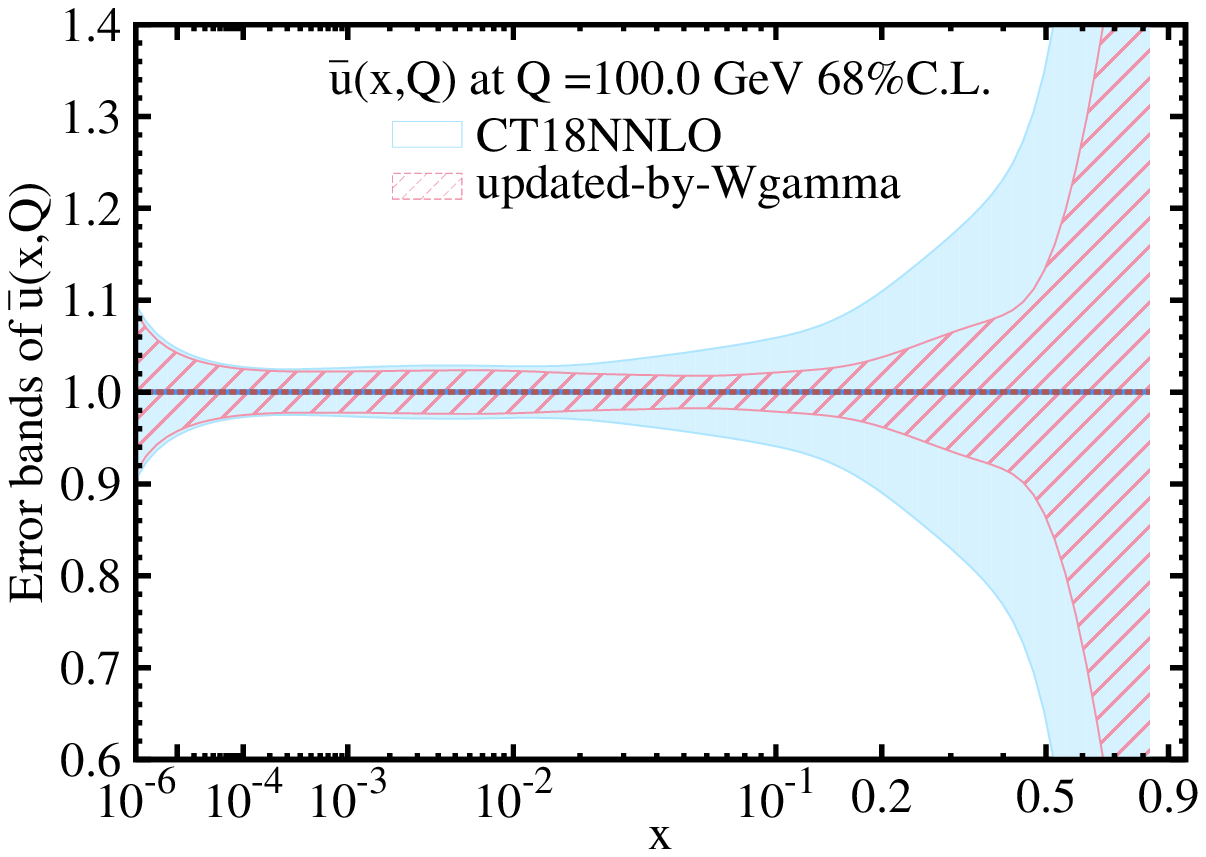}
\epsfig{scale=0.3, file=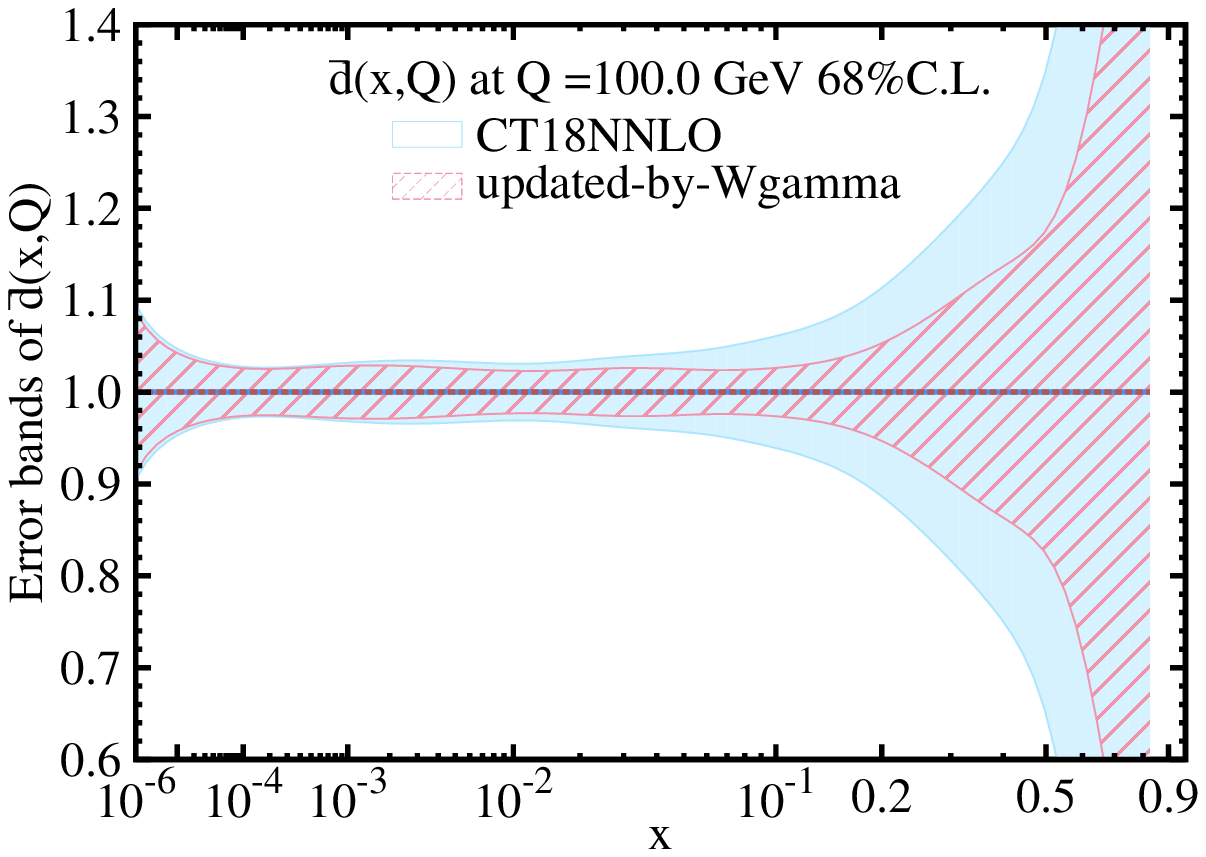}
\caption{\small Ratios of the central values and uncertainties to the CT18 central predictions of the 
valence $u$, valence $d$, 
$\bar{u}$ and $\bar{d}$ PDFs, before and after the PDF updating using $A^{W\gamma}_\text{boost}$. 
The blue band corresponds to the uncertainty before updating and the red band is after updating.}
\label{fig:UpdateWgamma}
\end{center}
\end{figure}

The results of using $A^{W\gamma}_\text{boost}$ 
to do the PDF updating are shown in Fig.~\ref{fig:UpdateWgamma}.  
With information of the quark exchanging fractions introduced, the uncertainties on the 
valence-$u$, valence-$d$, $\bar{u}$ and $\bar{d}$ PDFs are largely reduced. 
The proposed boost asymmetry contains unique information 
of the valence quarks in the small $x_S$ region and the 
sea quarks in the large $x_L$ region.
As depicted in the figures, the uncertainties of the sea $\bar{u}$ and 
$\bar{d}$ quark PDFs significantly reduced in the large $x_L$ region, while 
the valence $u$ and $d$ ones have better precision in the small $x_S$ region. 
Although 
$A^{W\gamma}_\text{boost}$ can also offer constraints on $q_i(x_L)\bar{q_j}(x_S)$, 
the single $Z$ and $W$ observations at the LHC 
would certainly cover such information with much larger statistics. 
For example, it was concluded in Ref.~\cite{WZconstrain} that the up 
and down valence quark PDFs can be better constrained in the large $x_L$ region by 
the high mass Drell Yan data. 
Therefore, 
the improvements on the small $x_S$ region antiquarks and 
the large $x_L$ region valence quarks are not significant by using the boost asymmetry.

\begin{figure}[!hbt]
\begin{center}
\epsfig{scale=0.3, file=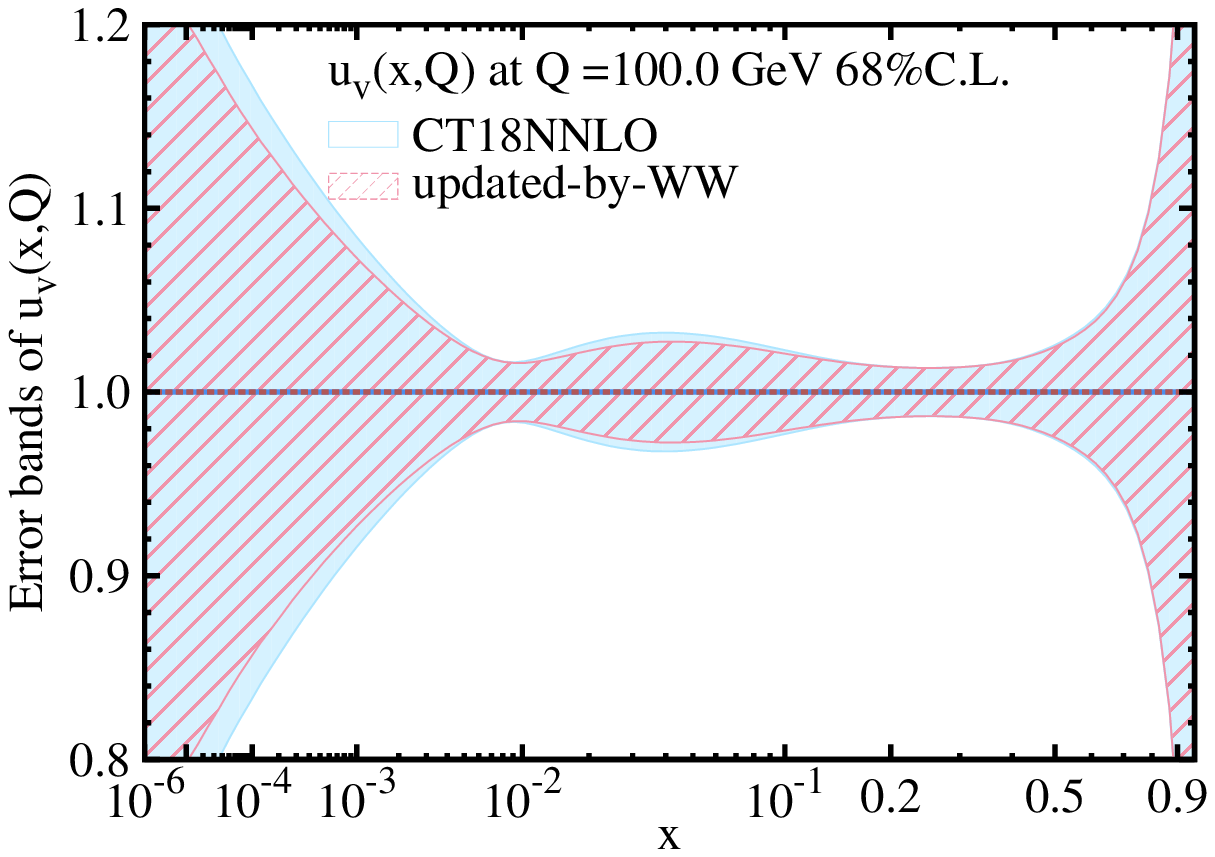}
\epsfig{scale=0.3, file=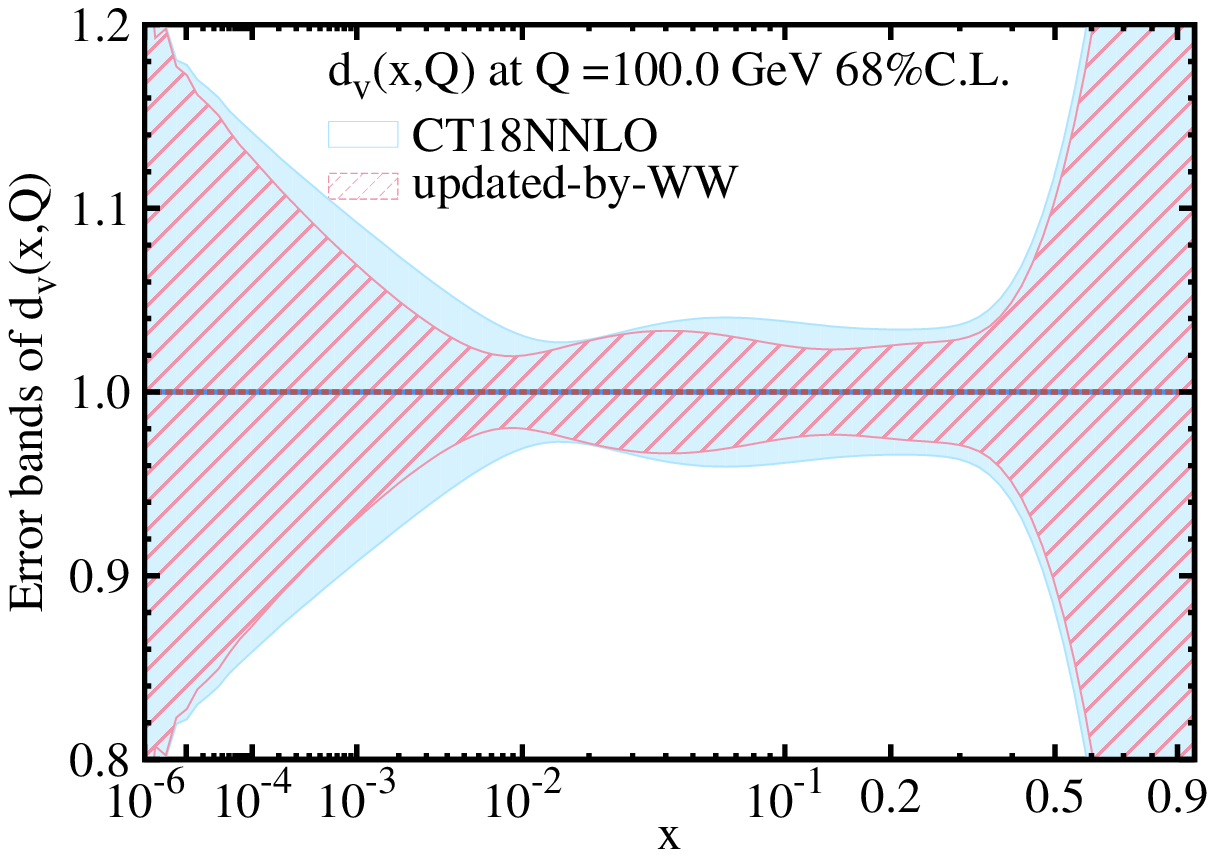}
\caption{\small Ratios of the central value and uncertainties to the 
CT18 central values of the valence $u$  
and valence $d$ PDFs, before and after the PDF updating using $A^{WW}_\text{boost}$. The blue 
band corresponds to the uncertainty before updating and the red band is after updating.}
\label{fig:UpdateWW}
\end{center}
\end{figure}

The impact of using $A^{WW}_\text{boost}$ is less significant. 
Firstly, the cross section 
of the $W^+W^-$ process is much smaller than the $W\gamma$ process. Secondly, 
unlike the $W\gamma$ measurement which can 
distinguish the $u\bar{d}$ and $d\bar{u}$ initial states, 
the $W^+W^-$ process cannot separate the $u\bar{u}$ 
from the $d\bar{d}$ ones. Thirdly, the sensitivity is 
of $A^{WW}_\text{boost}$ is further reduced 
due to the double neutrinos in both the $W$ boson decay.
Its leading effect, based on the 
CT18 updating, is on the valence $u$ and $d$ quark PDFs as shown in Fig.~\ref{fig:UpdateWW}. 
However, a potential impact of measuring $A^{WW}_\text{boost}$ 
is expected beyond its own sensitivity. 
As discussed in the introduction, the $A_{FB}$ observation in the single $Z$ 
Drell-Yan process at the LHC 
is believed to have high sensitivity on the quark exchanging fraction of $u\bar{u}$ and 
$d\bar{d}$, but not available in practice due to the strong correlation with $\effstw$. 
Reported in Ref.~\cite{AFBfactorization}, the $A_{FB}$ spectrum at 
the LHC has been analytically factorized with proton structure parameters 
representing the relevant parton information, 
so that the structure parameters can be 
determined together with $\effstw$ by simultaneous fit, 
with the correlation automatically taken into account. 
It is pointed out that the precision of the simultaneous fit is 
expected to be significantly improved if other data could be introduced in the fit, 
which ought to be $\effstw$-independent and providing the  
information on the quark exchanging fraction exactly same as $A_{FB}$ provides. 
The observable $A^{WW}_\text{boost}$ is an ideal input 
satisfying the requirement proposed in Ref.~\cite{AFBfactorization}, 
thus can be used to improve the precision both on the PDF and $\effstw$ by 
further reducing the correlation between them in the $A_{FB}$ measurement.

The detailed numbers given in the discussion could be 
different when higher order calculations and other PDFs are used in this test. However, 
the phenomenal conclusion that $A^{VV'}_\text{boost}$ can provide important information to the PDFs 
should be independent with the choice of event generators and PDFs. 
In this work, $A^{VV'}_\text{boost}$ is defined in terms of the boson and lepton rapidity. It could be 
defined with other kinematic variables such as boson and lepton energy under 
a specific experimental apparatus, if needed.

\section{Conclusion}

We propose the boost asymmetry in the $W\gamma$, $WZ$, and $W^+W^-$ events 
at the LHC 
as new experimental observables to constrain the PDFs. 
The kinematics of the two different bosons separately reflect the parton information 
of quarks and antiquarks in the initial state, respectively, and result 
in an asymmetry on the boosted boson. Such boost asymmetry can be used to constrain 
the quark exchanging fractions directly, which cannot be 
observed in the current single $W$ and $Z$ measurements. The observation on $A_\text{boost}$ 
in the $W\gamma$ and $WZ$ events is sensitive to the quark exchanging 
fraction of $u\bar{d}$ and $d\bar{u}$, while in the $WW$ events it is sensitive to 
that of $u\bar{u}$ and $d\bar{d}$ initial states. The $W^\pm\gamma$ events 
are particularly useful because the different strength of photon couplings to 
up and down quarks can further enhance the asymmetry. Impact study shows a reduction on 
the PDF uncertainties of relevant quarks when introducing $A_\text{boost}$ in to the 
PDF global analysis. The asymmetries would be helpful not only for PDF improvement, 
but also to other related topics such the measurement of $A_{FB}$ and the EW $\effstw$ parameter.

\section{Acknowledgements}
This work was supported by the National Natural Science Foundation of China under Grant No. 11721505, 11875245, 
12061141005 and 12105275, and supported by the ``USTC Research Funds of the Double First-Class Initiative''. 
This work was also supported by the U. S. National Science Foundation under Grant No. PHY-2013791. 
C.-P. Yuan is also grateful for the support from the Wu-Ki Tung endowed chair in particle physics.

\end{document}